\documentclass[twocolumn, prl]{revtex4}
\usepackage{graphicx}
\usepackage{epsfig}
\begin{document}
\title{Universal Thermometry for Quantum Simulation }
\author{Qi Zhou and Tin-Lun Ho}
\affiliation{Department of Physics, The Ohio State University, Columbus, OH 43210}
\date{\today}

\begin{abstract} 
Quantum simulation is a highly ambitious program in cold atom research currently being pursued in laboratories worldwide. The goal is to use cold atoms in optical lattice to simulate models for unsolved strongly correlated systems, so as to deduce their properties directly from experimental data.  An important step in this effort is to determine the temperature of the system, which is essential for deducing all thermodynamic functions.   This step, however, remains difficult for lattice systems at the moment.  
Here, we propose a method  based on a generalized fluctuation-dissipation theorem. 
It does not reply on numerical simulations and is a universal thermometry for all quantum gases systems including mixtures and spinor gases. It is also unaffected by photon shot noise.
 \end{abstract}

\maketitle

At present, there is worldwide experimental effort to simulate theoretical models for strongly correlated quantum systems using cold atoms in optical lattices.  If successful, these simulations can provide detailed thermodynamic information for  many models whose solutions are unknown, even some of them (such as 2D Hubbard model) have been studied for decades.  To deduce the  thermodynamic properties of these models directly from experiments,  it is necessary to determine three quantities accurately : density $n$, chemical potential $\mu$, and temperature $T$\cite{HoZhou1, Nandini}. 
The recent experiment of Cheng Chin's group\cite{Cheng}  using in situ density profile to identify directly  the thermodynamic phases for boson Hubbard systems is a very important step  toward realizing the full power of quantum simulation\cite{HoZhou1}.  The prospect of this realization is further enhanced by the impressive improvement  in resolution of density imaging recently developed in Markus Griener's group\cite{Greiner}.  The next crucial step is to have an accurate temperature determination. 

Often, the temperature of a lattice gas is estimated by assuming the lattice is turned on adiabatically. One then equate the entropy of the final state $S_{f}(T_{f})$ to that of the initial state $S_{i}(T_{i})$, (i.e. the state before the lattice is switched on),  and then deduce the final temperature $T_{f}$ from the initial temperature $T_{i}$ through this relation. One factor detrimental to this procedure is  the intrinsic heating caused by spontaneous emission, which occurs  as the lattice is turned on, and during the time when experiment is  performed\cite{Troyer}.  To make things worst, the 
entropy function $S_{f}(T)$ of many systems of interest remains unknown.  So the errors of this method are uncontrolled\cite{comment0}.  

For quantum gases in a single trap without optical lattice, their temperatures can be deduced from the density profile at the surface, which has the Boltzmann form. In principle, one can apply the same method for lattice quantum gases, 
as interaction effects becomes unimportant near the surface.  However, an accurate determination of the density profile near the surface will require improving the imaging resolution to a single site. It will also require repeating the experiment many times so as to achieve a good signal to noise ratio. To avoid these demands, many experiments resort to the aforementioned adiabatic assumption for temperature determination.  However, due to the uncontrolled errors in this method,  it is desirable to have an alternative scheme which is robust and free of all the problems mentioned above. We also note that by studying the density at the surface,  one can not determine whether the entire sample is in global equilibrium. 

In this paper, we present a new scheme to determine the temperature of trapped quantum gases based on the fluctuation-dissipation theorem {\em for non-uniform systems}. This method applies to all quantum gas systems (single component gases, mixtures, spinor gases) and is unaffected by  background photon shot noise. It can also be used to deduce magnetic susceptibility of bulk systems. This method does not require numerical input, and can tell whether the system is in global equilibrium.

{\bf A1. The proposal:} We begin with two basic assumptions used in most experiments on quantum gases which  have been justified in many cases.  The first is that the density $n({\bf r})$ of a quantum gas in a trap $\hat{V}({\bf r})$ can be calculated in grand canonical ensemble , i.e.  $n({\bf r}) = n({\bf r};  T, \mu) $, where
\begin{equation}
n({\bf r};  T, \mu) = \frac{ {\rm Tr}  \hat{n}({\bf r}) e^{-\beta(\hat{H}+ \hat{V} - \mu \hat{N}) } }{ {\rm Tr}  e^{-\beta(\hat{H}+ \hat{V} - \mu \hat{N}) } } \equiv \langle \hat{n}({\bf r}) \rangle_{T, \mu}. 
\label{gc} \end{equation}
where $\beta = 1/(k_{B}T)$, $\hat{H}$ is the Hamiltonian without trapping potential, $T$ is the temperature and $\mu$ is the chemical potential. 
The second is that $n({\bf r};  T, \mu)$ is given accurately by local density approximation (LDA), i.e.  \begin{equation} 
n({\bf r};  T, \mu) = n_{o}(\mu({\bf r}), T), \,\,\,\,\,\,\, \mu({\bf r}) = \mu - V({\bf r}), 
\label{LDA} \end{equation}
where $ n_{o}(\nu, T)$ is the density of a homogeneous system with hamiltonian $\hat{H}$ and chemical potential $\nu$, i.e. 
$n_{o}(\nu, T) = { {\rm Tr} e^{-\beta(\hat{H}-\nu \hat{N})}  \hat{N}  }/({\Omega} {\rm Tr}  e^{-\beta(\hat{H}- \nu \hat{N})} )$, and $\Omega$ is the volume of the homogenous system.  For lattice quantum gases, LDA is justified if the variation of  trapping potential between neighboring sites is small compared with the hopping matrix element. 
Eq.(\ref{gc}) implies 
\begin{equation}
k_{B}T \frac{\partial \langle \hat{n}({\bf r})\rangle  }{\partial \mu} =  \int {\rm d}{\bf r'} \left[  \langle \hat{n}({\bf r})  \hat{n}({\bf r'}) \rangle - 
 \langle \hat{n}({\bf r})  \rangle  \langle \hat{n}({\bf r'})  \rangle \right], 
\label{fl0}    \end{equation}
 where $\langle ...\rangle = \langle ...\rangle_{T, \mu}$. 
For an isotropic harmonic trap $V({\bf r}) = \frac{1}{2}M \omega^2 {\bf r}^2$ with frequency $\omega$, Eq.(\ref{LDA}) becomes
\begin{equation}
-\frac{k_{B}T}{M\omega^2 r}  \frac{\partial \langle  \hat{n}({\bf r})\rangle  }{\partial r} = \int {\rm d}{\bf r'}  \left[  \langle \hat{n}({\bf r})  \hat{n}({\bf r'}) \rangle - 
 \langle \hat{n}({\bf r})  \rangle  \langle \hat{n}({\bf r'})  \rangle \right], 
\label{fl1} \end{equation}
 or simply
 \begin{equation}
-\frac{k_BT}{M\omega^2 r}  \frac{\partial \langle  \hat{n}({\bf r})\rangle  }{\partial r} = \langle \hat{n}({\bf r}) \hat{N} \rangle - 
 \langle \hat{n}({\bf r})  \rangle  \langle \hat{N}\rangle. 
\label{fl2} \end{equation}

Eq.(\ref{fl2}) suggests a convenient way to determine temperature.  Suppose we repeat the experiments $Q$ times, and label the measured quantities of each sample by a superscript ``$i$",  $i=1,2,3, ...Q$. Let  $n^{(i)}({\bf r})$ be the density profile of the $i$-th sample,  and $N^{(i)}= \int n^{(i)}({\bf r})$ be 
the total number of particle of that sample.  The averages of these quantities over all $Q$ samples will be denoted as  $\overline{n({\bf r})}$ and   $\overline{N}$, where
$\overline{x} \equiv \sum_{i=1}^{Q} x^{(i)}/Q$. 
In the limit where $Q>>1$, Eq.(\ref{fl2}) can be written as $L({\bf r})= R({\bf r})$,  where 
$R({\bf r})  =  \overline{ n({\bf r}) N} - \overline{n({\bf r})}\,\, \overline{N}$, or
\begin{equation}
R({\bf r}) 
= Q^{-1} \sum_{i= 1}^{Q}  n^{(i)}({\bf r})  N^{(i)}+
Q^{-2}  \sum_{i,j= 1}^{Q}   n^{(i)}({\bf r})  N^{(j)}; 
\end{equation}
\begin{equation}
L({\bf r})  = - Q^{-1} \sum_{i= 1}^{Q}  \frac{k_{B}T^{(i)}}{M\omega^{2} r} \frac{\partial n^{(i)}({\bf r})}{\partial r}.
  \hspace{0.73in}
\label{LL}  \end{equation}
That we label temperature $T$ with a superscript $i$ is because in real experiments, there are fluctuations in temperature in each of the samples due to the initial evaporation process.  In the following, we shall assume the temperature fluctuations from sample to sample are sufficiently small compared to the mean temperature so that they can be ignored. In this case, we can set $T^{(i)}$ to its mean values, which we simply denote as $T$, and Eq.(\ref{LL}) becomes 
\begin{equation} 
L({\bf r}) = T{\cal L}({\bf r}), \,\,\,\,\,\, {\cal L}({\bf r}) =  - \frac{k_{B}}{M\omega^{2} r} \frac{\partial \overline{n{(\bf r})}}{\partial r}. 
\label{first}   \end{equation}
Eq.(\ref{fl2}) them implies $T= R({\bf r})/{\cal L}({\bf r})$ at any position ${\bf r}$. 
 
There is, of course, the practical matter of how many samples is needed to average over to reach the thermal average.  A very large value of $Q$ will not be practical.  To achieve fast convergence, one can suppress the noise by averaging over a ring  of thickness $\epsilon$, resulting in the function (say, in the 2D  case) 
$\zeta(\rho) =  \int^{\rho+\epsilon}_{\rho} \eta(\rho') \rho' {\rm d}\rho' /\Omega(\rho)$, where $ \eta(\rho) = \int^{2\pi}_{0} {\rm d} \theta \, n(\rho, \theta)$  is the angle integrated density at radius $\rho$,  $\Omega(\rho)=\pi [(\rho+\epsilon)^2-\rho^2]$ is the area of the ring being averaged over,  and $(\rho,\theta)$ are polar coordinates.
From Eq.(\ref{fl2}), we obtain temperature $T$ as 
\begin{equation} 
T= {\cal R}(\rho)/{\cal L} (\rho), 
\label{T}\end{equation}
\begin{equation} 
 {\cal R}(\rho) = \overline{ \zeta (\rho) N} - \overline{\zeta(\rho)}\, \overline{N}, 
\label{R} \end{equation}
 \begin{equation}
 {\cal L}(\rho) = - \left( \frac{k_{B}}{M\omega^2\Omega(\rho)}\right)  \int^{\rho+\epsilon}_{\rho} {\rm d}s
 \overline{ \frac{{\rm d}\eta(s)}{{\rm d}s} }. 
\label{L}\end{equation}
Eq.(\ref{T}) holds for all radii $\rho$. 
In the 3D case, the quantity easily accessible is column integrated density. 
The corresponding expression for Eq.(\ref{T}) is to replace  $\zeta(\rho)$ and $\eta(\rho)$  by their column integrated analogs $a(\rho)$ and $b(\rho)$, where $a(\rho) = \int^{\rho+\epsilon}_\rho \rho' b(\rho') {\rm d} \rho'/\Omega(\rho)$, $b(\rho)=  \int^{2\pi}_{0} {\rm d}\theta \int {d z}\,\, n(\rho,\theta, z)$, and $(\rho, \theta, z)$ are cylindrical coordinates. 

\begin{figure}[tbp]
\begin{center}
\includegraphics[width=2.2in]{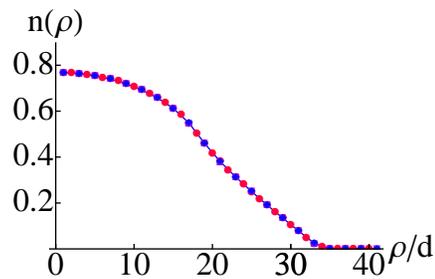}
\end{center}
\caption{Density profile in the trap. Red dots: Ensemble averaging over 2000 configurations.
Blue diamonds:LDA. Purple boxes: Exact density in the trap.}\label{fig: fig1}
\end{figure}

To illustrate the working of Eq.(\ref{T}), we consider a 2D  ideal Fermi gas in a square lattice 
with Hamiltonian $\hat{H}= - t\sum_{\bf \langle R, R' \rangle, \sigma}c^{\dagger}_{\bf R, \sigma}c^{}_{\bf R', \sigma}$ and in an overall harmonic potential $\hat{V} = \frac{1}{2} \sum_{\bf R} M \omega^2 {\bf R}^2 c^{\dagger}_{\bf R}c^{}_{\bf R}$ with frequency $\omega$. Here, $t$ is the hopping matrix element, $ \langle R, R' \rangle$ means neighboring sites, and $c^{\dagger}_{\bf R, \sigma}$ creates a fermion at site ${\bf R}$ with spin $\sigma$. 
The equilibrium density of this non-uniform system is $\langle \hat{n}({\bf r})\rangle = \sum_{\alpha} |u_{\alpha}({\bf r})|^2 f( E_{\bf \alpha}) $, where $f(x)= (e^{(x-\mu)/k_{B}T} +1)^{-1}$ is the Fermi distribution function,  $E_{\alpha}$ and $u_{\alpha}({\bf r})$ are eigen-energies and eigen-functions of the system $H+V$.

 In Figure 1, we show the equilibrium density of a system with temperature $T/t= 0.1$ and a chemical potential $\mu$ adjusted so that the number of particles is  $N=1200$. We also show on the same plot the  LDA result, which differs from the grand canonical result by less than $0.1\%$ and is invisible in the figure, justifying the assumptions we mentioned at the beginning. 
 To generate an equilibrium ensemble, we start with an arbitrary assignment of 0 and 1 of the  occupation numbers $\{ n_{\alpha} \}$ of the energy levels $\{ E_{\alpha} \}$ up to a very large cutoff $\Lambda$, and evolve the set $\{ n_{\alpha} \}$ with Monte Carlo scheme. We have generated 2000 configurations $\{ n_{\alpha} \}$ after the system has reached equilibrium, which we refer to as the equilibrium ensemble.  The average of the occupation number $n_{\alpha}$ is given by the Fermi distribution (with $0.1\%$ accuracy in our calculation), and that there are no correlations between the occupations of different energy levels (which is the properties of ideal gas). 
  \begin{figure}[tbp]
\includegraphics[width=1.65in]{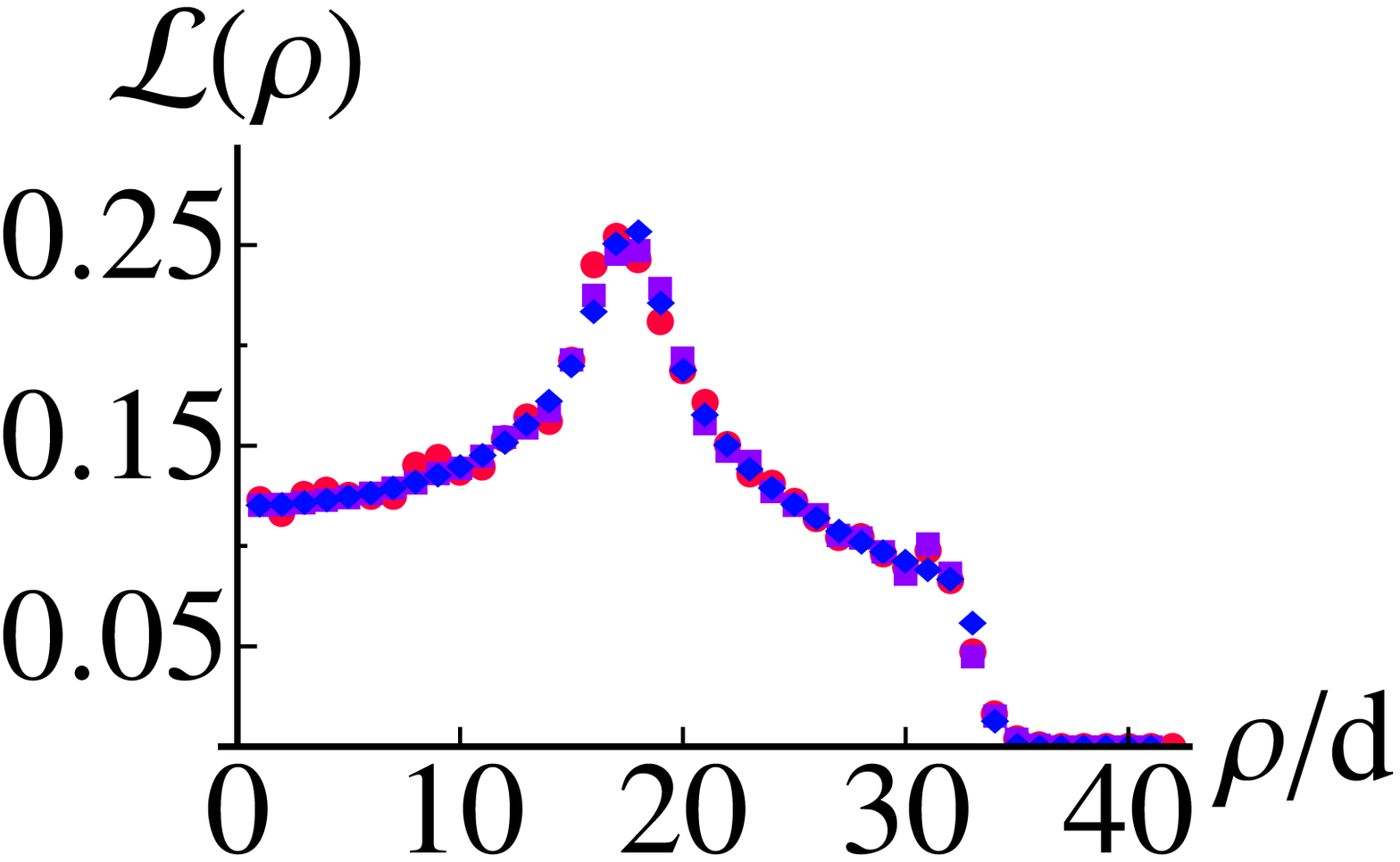}
\includegraphics[width=1.65in]{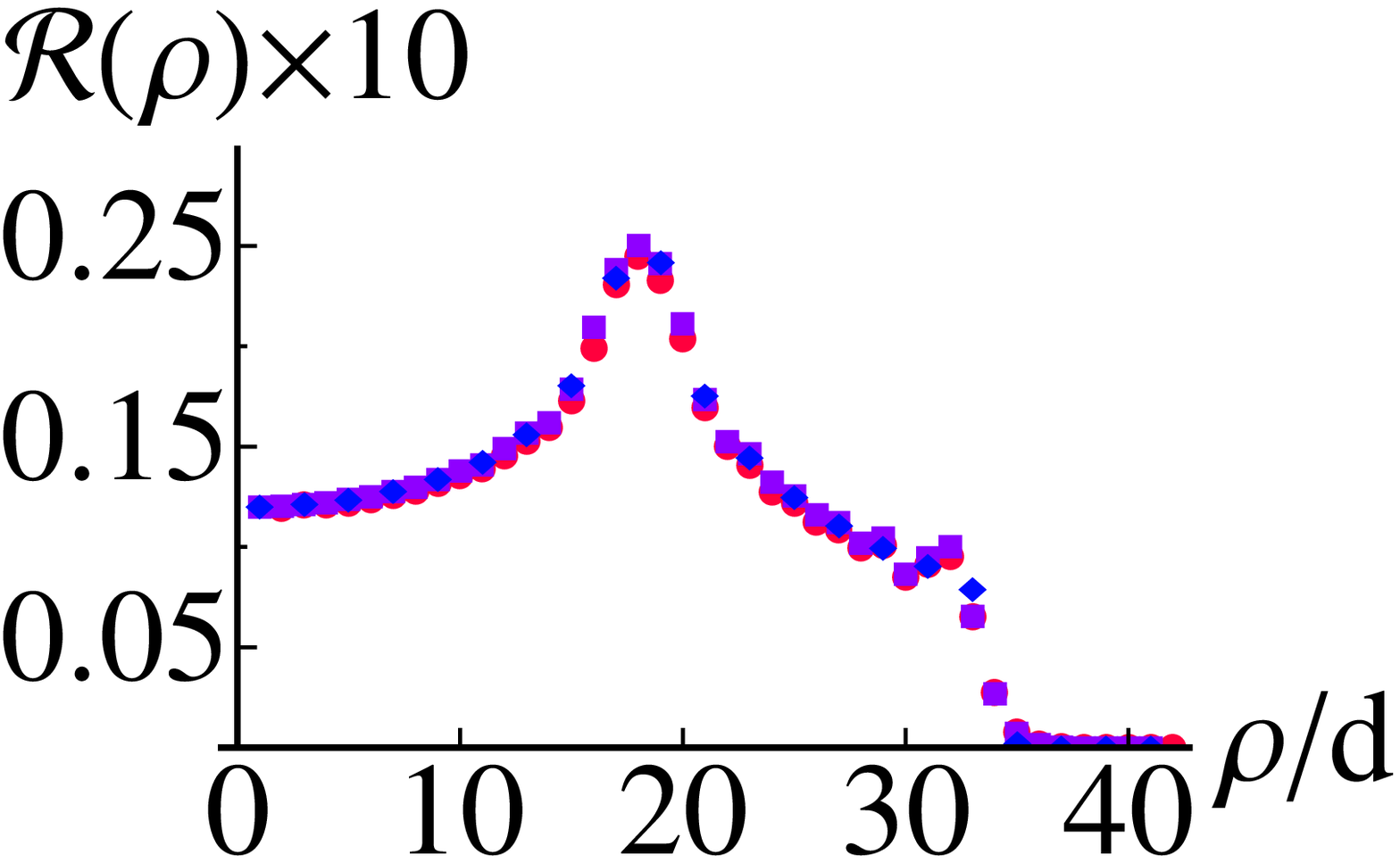}
\caption{Compressibility (left) and number fluctuations (right) in the trap. Red dots: averaging over 50 configurations.  Purple boxes: Exact results in the trap. Blue diamonds: LDA. }\label{fig: fig2}
\end{figure}
Within this equilibrium ensemble, we randomly select $Q$ configurations, which would correspond to $Q$ measured samples in experiments.  The level occupation of these configurations will be labeled as 
 $\{ n^{(i)}_{\alpha}\}$, $i=1$ to $Q$. 
 We then evaluated the densities  $\overline{n({\bf r})} = \sum_{\alpha} |u_{\alpha}({\bf r})|^2 \overline{n_{\bf \alpha}} $ and the number fluctuations $\overline{n({\bf r})N}-\overline{n({\bf r})} \,\,\overline{N}=  \sum_{\alpha} |u_{\alpha}({\bf r})|^2(\overline{n_\alpha^2}-\overline{n_\alpha}^2) $ of these $Q$ samples, and  the angle averaged 
 quantities  ${\cal R}(\rho)$ and ${\cal L}(\rho)$ defined in Eq.(\ref{R}) and (\ref{L}), where $\overline{n_\alpha}=Q^{-1}\sum_{i=1}^{i=Q}n_\alpha^{(i)}$ and $\overline{n_\alpha^2}=Q^{-1}\sum_{i=1}^{i=Q}n_\alpha^{(i)2}$.  The dependence of these quantities as a function of raduis $\rho$ for the case of $Q=50$ is shown in Figure 2\cite{VanHove}.  In this figure, we also include the LDA result, which differ from the data only by a few percent. This shows once again that LDA is an excellent approximation. Figure 3 displays the pairs  $({\cal R}(\rho), {\cal L}(\rho) )$ in the ${\cal L}-{\cal R}$ plot.  According to Eq.(\ref{T}), all the points should fall in a straight line with slope given by $T$. By randomly choosing 50 samples in the equilibrium ensemble, we obtain a temperature within  $3\%$ of the actual temperature. If we increase the number of sample to be averaged to $Q=200$, the accuracy in temperature increases to 1$\%$.  
 We have repeated our calculation for the same system at lower temperature $T/t=0.02$ and have found the same accuracy in temperature determination. 
  
We stress that the angular average is crucial for our scheme. Due to the self-averaging property of the equilibrium ensemble, the angular average enhances signal to noise significantly, and amounts to a significant increase in the number of configurations averaged. The high accuracy of temperature determination by averaging only 50 samples makes our scheme practical. 
We would also like to point out that if the system is not in global equilibrium, but was able to establish different temperatures in different parts of the sample, then the points in the ${\cal R}$-${\cal L}$ plot will fall into a few straight lines with different slopes.

{\bf A2. Fluctuations in $\mu$ and $T$:} 
Next, we consider the effect of  fluctuations in $T$. We have generated density profiles of  equilibrium ensembles at different temperatures while keeping $\mu$ and $\omega$ fixed.  We find that with $1\%$ ($5\%$) temperature fluctuations,  the accuracy for temperature determination after averaging 50 samples remains at $5\%$ (changed to between $5\%$ to 10$\%$). We have repeated our calculations for similar fluctuations in $\mu$ (which amounts up to $5\%$ fluctuations in $N$), and have found the results. This shows our scheme is robust against these fluctuations, and that Eq.(\ref{first}) is justified. 

{\bf  A3. Local density fluctuation : }  We again consider the 2D case. In experimental analysis,  one divides up the real space into  a square lattice of units cells (i.e. bins) and count particle number $n_{\bf R}$ where ${\bf R}$ labels the location of the bin.  Eq.(\ref{fl0}) then becomes 
\begin{equation}
T C_{\bf R} = D_{\bf R} +  F_{\bf R}, 
\label{short} \end{equation}
where $C_{\bf R}$ and $D_{\bf R}$ are the local compressibility and local density fluctuation, 
\begin{equation}
C_{\bf R} = - \frac{k_{B}}{M\omega^2 r}\frac{\partial \langle n_{\bf R}^{}\rangle }{\partial r}, \,\,\,\,\,\,\,
D_{\bf R} =
\langle n_{\bf R}^{2} \rangle - \langle n_{\bf R} \rangle^{2},  
\end{equation}
 $F_{\bf R}  = \sum_{\bf R'\neq {\bf R}} G_{\bf R, R'} $ is  the fluctuation  due to neighboring bins, and 
 $G_{\bf R, R'} =  \langle n_{\bf R}^{} n_{\bf R'}^{}\rangle -  \langle n_{\bf R}^{}\rangle
 \langle n_{\bf R'}^{}\rangle$ 
 is the density correlation at different bins. 
If the range of $G_{\bf R, R'}$  at ${\bf R}$ happens to be very short, which may occur if the system at ${\bf R}$ is in the Mott phase or the normal phase, then $F_{\bf R}\sim 0$ and  Eq.(\ref{short}) implies that  temperature is simply the ratio $T=D_{\bf R}/C_{\bf R}$. Thus, if we plot $D_{R}= \int_{\theta} D_{\bf R}$ against $C_{R}=\int_{\theta} C_{\bf R}$, where $\int_{\theta}$ denotes angular average, we will find many points  $(C_R, D_R)$ in the $C-D$ plot fall onto a straight line while many other points do not. The former comes from the regions of $\{ R \}$ with short range density correlations, while the latter from regions with longer range correlations. This suggests a simple way to use local density fluctuation to determine temperature: Even though only a portion of the curve $(C_R, D_R)$ fall on a straight line, one can still determine $T$ from the slope of this straight line\cite{comment}.  
This method, while convenient, is not as general as that discussed in ${\bf A.1}$,  for it requires a significant part of the system to have short range density correlations. 


{\bf A4. A further simplification:}  Finally, we note that by integrating overall ${\bf r}$, and using the fact that ${\rm d}\mu = -\frac{1}{2} M \omega^2 {\rm d}r^2$, Eq.(\ref{fl0}) becomes (in the $2D$ case)
\begin{equation}
\frac{2\pi k_{B} T n_{\bf R=0}^{} }{M\omega^2} = \langle N^2\rangle - \langle N\rangle^2. 
\label{simple} \end{equation}
One can therefore also determine $T$ from the central density and total number fluctuation. 
In the 3D case, $n_{\bf R=0}$ will be replaced by the column density at ${\bf R=0}$. 
Despite the simplicity of Eq.(\ref{simple}), the algorithm discussed in ${\bf A1}$ remains more robust, as $T$ is determined by the contributions of all radii $\rho$. 

  \begin{figure}[tbp]
\begin{center}
\includegraphics[width=2.8in]{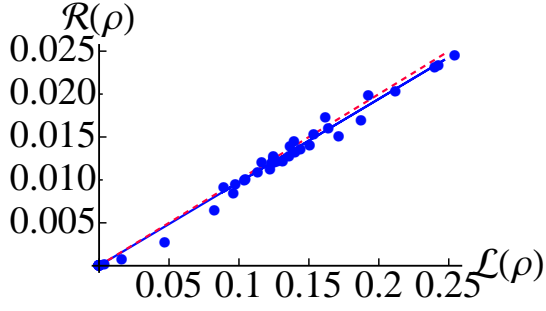}
\end{center}
\caption{Linear fit for $\{\cal{L}(\rho), \cal{R}(\rho)\}$ to extract $T$. Blue dots are the results of $\cal{L}(\rho)$ and $\cal{R}(\rho)$ from averaging 50 configurations. Blue straight line is the fitting results. Red straight dashed line represents the real temperature $T=0.1t$.}\label{fig: fig3}
\end{figure}

{\bf B. Photon shot noise}:  In real imaging process, there is photon shot noise. The measured atom number  (denoted as $\hat{n}^{ex}({\bf r})$) is related to the the actual  atom number $\hat{n}({\bf r})$ as  
$\hat{n}^{ex}({\bf r}) = \hat{n}({\bf r}) + \hat{\nu}({\bf r})$, 
where $\hat{\nu}({\bf r})$ is the contribution due to  photon shot noise. Since photon shot noise is a property of the laser, its probability distribution ${\cal P}_{p}[\nu]$ is independent  from that of the density distribution,  ${\cal P}_{a}[n]$, which is given by thermodynamics. Averaging over different equilibrium samples (denoted as $\langle ..\rangle_{a,p}$) means averaging over these two independent distributions.  In other words, the experimentally measured density  $n^{ex}({\bf r}) = \langle \hat{n}^{ex}({\bf r})\rangle_{a,p}  $ is 
\begin{equation}
n^{ex}({\bf r}) = n({\bf r}) + c, \,\,\,\,\, \,n({\bf r}) =\langle \hat{n}({\bf r})\rangle_{a},  \,\,\,\, c= \langle \hat{\nu}({\bf r}) \rangle_{p}, 
\end{equation}
where $c$ the average of background photon shot noise,  which can be calibrated by taking images in the absence of atoms. We shall also assume the short noise has no significant spatial correlation, i.e. 
\begin{equation}
\langle \nu({\bf r}) \nu({\bf r'})\rangle_{a} = c^2 + c_{1} \delta({\bf r}-{\bf r'}), 
\label{nunu} \end{equation}
where $c_{1}$ is shot noise fluctuation about its mean $c$. 

Since the average noise $c$ is independent of $\mu$, we have 
\begin{equation}
k_BT\frac{\partial n^{ex}({\bf r})}{ \partial \mu} = k_BT\frac{\partial n({\bf r})}{\partial \mu} = \langle \hat{n}({\bf r}) \hat{N}\rangle_{a} - n({\bf r}) N, 
\label{dnex} \end{equation}
$N = \langle \hat{N}\rangle_{a} =  \int {\rm d}{\bf r} \, n({\bf r})$. 
Next, consider the fluctuations of measured density, 
$R^{ex}({\bf r}) = \langle \hat{n}^{ex}({\bf r}) \hat{N}^{ex} \rangle_{a,p}$ $-\langle \hat{n}^{ex}({\bf r}) \rangle_{a,p} \langle \hat{N}^{ex} \rangle_{a,p} $. 
We note that  
\begin{eqnarray}
\langle \hat{n}^{ex}({\bf r}) \hat{N}^{ex} \rangle_{a,p} = \langle ( \hat{n}({\bf r}) + \hat{\nu}({\bf r}) ) ( \hat{N} + \int {\rm d} {\bf r'} \hat{\nu}({\bf r'})) \rangle_{a,p} \,\,\,\, \\
= \langle \hat{n}({\bf r})\hat{N}\rangle_{a}  + c N + n({\bf r}) V c +  \int {\rm d} {\bf r'} \langle \hat{\nu}({\bf r})\hat{\nu}({\bf r'})) \rangle_{p}  \,\,\,\,\,\,
 \label{nN} \end{eqnarray}
 where $V$ is the volume for photo collection; and 
\begin{eqnarray}
\langle \hat{n}^{ex}({\bf r}) \rangle_{a,p} \langle \hat{N}^{ex} \rangle_{a,p} 
= (n({\bf r}) +c) ( N + Vc)  \,\,\,\, \\
= n({\bf r})N + cN + n({\bf r})Vc + Vc^2.  \hspace{0.5in}
\label{n-N-} \end{eqnarray}
Subtracting Eq.(\ref{n-N-}) from (\ref{nN}), and using Eq.(\ref{nunu}), we have
\begin{equation}
R^{ex} ({\bf r}) =   \langle \hat{n}({\bf r}) \hat{N}\rangle_{a} - n({\bf r}) N + c_{1} .
\label{Rex}  \end{equation}
Eq.(\ref{dnex}) and (\ref{Rex}) then imply
\begin{equation}
\frac{\partial n^{ex}({\bf r})}{ \partial \mu} = \langle \hat{n}^{ex}({\bf r}) \hat{N}^{ex} \rangle_{a,p} -  
\langle \hat{n}^{ex}({\bf r}) \rangle_{a,p} \langle \hat{N}^{ex} \rangle_{a,p}  - c_{1}. 
\end{equation}

Eq.(\ref{T}) then becomes 
\begin{equation}
T = - \left( \frac{M\omega\Omega(\rho)}{k_{B}}\right) \frac{ \overline{ \zeta^{ex}(\rho) N} - \overline{\zeta^{ex}(\rho)}  \,\,\, \overline{N} -{c}_{1}
 }{\int_\rho^{\rho+\epsilon}ds d\overline{ \eta^{ex}(s)}/ds}.
\label{TT} \end{equation} 
In 3D, $\zeta^{ex}({\rho})$ , $\eta^{ex}({\rho})$ and $c_1$ shall be replaced by $a^{ex}(\rho)=\int^{\rho+\epsilon}_\rho{\rm d} \rho' \rho' \int^{2\pi}_{0} {\rm d}\theta \int {d z}\,\, n^{ex}(\rho',\theta, z) /\Omega(\rho)$ and $b^{ex}(\rho)=\int^{2\pi}_{0} {\rm d}\theta \int {d z}\,\, n^{ex}(\rho,\theta, z)$ and ${\cal C}_{1}=\int {\rm d}z \, c_{1}$ respectively. 

{\bf Conclusion:}  We have shown that density fluctuation is a powerful way to determine the temperature of a trapped gas. It is clear from our derivation that this method applies to other systems such as  mixtures and spinor gases. The fact that the temperature can be determined by the fluctuation at every point in the sample provides considerable cross checks on the accuracy of the result.  Our method can also reveal situations where different regions of the sample  are in equilibrium within themselves but not with each other. At present, all methods of thermometry requires the input of specific theoretical modeling.  Our method replies only on thermodynamics. It is therefore immune from errors of theoretical modeling, and is in line with the true spirit of quantum simulation, i.e. finding information of unsolved models without specific theoretical input. 

We thank Cheng Chin for very stimulating discussions. This work is supported by NSF Grants DMR0705989, PHY05555576,
 and by DARPA under the Army Research Office Grant Nos. W911NF-07-1-0464, W911NF0710576.


\begin{thebibliography}{99}
\bibitem{comment0} Although there are cases where $S(T)$ can be calculated numerically, such theoretical input defeats the purpose of quantum simulation, which is to obtain physical properties including $S(T)$ directly from experimental data.
\bibitem{comment} From Fig. 3 and Fig.4  in ref.\cite{Cheng}, which plots ($C_R$ vs $n_R$) and ($D_R$ vs $n_R$) respectively, one can see that a straight line portion emerges in the ($C_R$-$D_R$) plot.
\end{thebibliography}
\end{document}